\documentclass[preprint2]{aastex}

\newcommand{\ie}{{\it i.e.\/}}

\newcommand{\myemail}{mledlow@gemini.edu}
\newcommand{\kms}{km sec$^{-1}$\/}

\shortauthors{Ledlow et al.}
\shorttitle{Large-Scale Jet in a Spiral Galaxy} 

\begin{document}

\title{A Large-Scale Jet and FR I Radio Source in a Spiral Galaxy: The
  Host Properties and External Environment}


\author{Michael J. Ledlow\altaffilmark{1}}
\affil{Gemini Observatory, AURA, Casilla 603, La Serena, Chile} 
\email{\myemail}

\author{Frazer N. Owen, Min S. Yun\altaffilmark{2}}
\affil{National Radio Astronomy Observatory\altaffilmark{3}, Socorro, NM 87801} 
\email{fowen@pilabo.aoc.nrao.edu, myun@astro.umass.edu}

\and 

\author{John M. Hill}
\affil{Steward Observatory, University of Arizona, Tucson, Arizona, 85721} 
\email{jhill@as.arizona.edu}

\altaffiltext{1}{Also: Institute for Astrophysics, University of New
  Mexico, Albuquerque, NM 87131}

\altaffiltext{2}{Present Address: Astronomy Department, University of
  Massachusetts, Amherst, MA 01003}

\altaffiltext{3}{The National Radio Astronomy Observatory is operated
  by Associated Universities, Inc., under contract with the National
  Science Foundation.}

\begin{abstract}
  We have identified a large ($\approx 200h_{75}^{-1}$ kpc), powerful
  double radio source whose host galaxy is clearly a disk and most
  likely a spiral.  This FR I-like radio galaxy is located very near
  the center of the Richness Class 0 cluster Abell 428.  The existence
  of such an object violates a fundamental paradigm for radio loud
  AGN's.  In paper I (Ledlow,Owen, \& Keel,1998,\apj,495,227), we
  showed that this object was most consistent with a spiral host
  classification with optical emission-line ratios and colors
  suggestive of an active nucleus. However, we were not able to
  confirm actual radio jet emission based on the maps available at
  that time.  In this paper, we present new, higher resolution radio
  imaging, a radio/mm continuum spectrum for the nucleus, a detection
  of {\sl HI} absorption against the bright radio core, an upper-limit
  to {\sl CO} emission and the gas mass, and 70 (68 new) optical
  redshifts measured in the direction of Abell 428.  We confirm the
  existence of a radio jet at 20cm, extending 42$h_{75}^{-1}$ into the
  southern lobe.  At 3.6cm, we also detect a nuclear jet similar in
  length to that in M87, although 10 times weaker.  We believe that
  this is the first detection of a radio jet on these scales in a
  disk/spiral host galaxy. The nuclear radio spectrum is similar to
  many blazar or QSO like objects, suggesting that the galaxy harbors
  an imbedded and obscured AGN. We model a turnover in the spectrum at
  low frequencies as a result of Free-Free absorption.  We detect very
  strong and narrow {\sl HI} absorption, with nearly the entire 20 cm
  continuum flux of the core being absorbed, implying an unusually
  large optical depth ($\tau \approx 1$). The most consistent model is
  that we are viewing the nucleus through a disk-like distribution of
  gas in the ISM, possibly through a spiral arm or a warp to account
  for the above average column density.  From the radial velocity
  distribution, we find that Abell 428 is in fact made up of at least
  2 clumps of galaxies separated by $\sim 3300$ \kms, which themselves
  appear to be imbedded in a nearly continuous distribution of
  galaxies over 13000 \kms in velocity space.  Thus, the environment
  around this unusual radio source is more like that of a poor galaxy
  group imbedded in a filament-like structure viewed end-on.
\end{abstract}



\section{Introduction}

It has been a standard paradigm and accepted fact for some time that
giant double-lobed, jet-fed radio sources (Fanaroff \& Riley types I
or II; \citet{fr74}) are hosted by elliptical galaxies or distorted
versions thereof (\ie such as merger remnants).  The reasons
for such a strong correlation between the AGN phenomena and galaxy
type has been interpreted as telling us something about the central
engine and black holes in different hosts.  Alternatively (or
additionally), the strong morphological correlation may be related to
the fragile jets being disrupted by the dense, rich ISM in spiral
galaxies.  The fact that nuclear jets are a common feature in Seyfert
galaxies is a strong argument for the latter \citep{baum93}.

While there have been a number of possible counterexamples to the
elliptical/spiral dichotomy, they have all turned out to be either
source misidentifications or gas-rich merger products more consistent
with an elliptical or bulge-dominated host to the imbedded AGN (such
as 3C305, \citet{heckman85}).  However, in Paper I \citep{paper1}, we
presented the discovery of 0313-192 in Abell 428, which appears to be
the first confirmed case of a powerful ($\log P_{20cm}=23.95$ W
Hz$^{-1}$) double radio source in a clearly disk-dominated host. This
object was found from a radio survey of more than 500 Abell clusters
(\citet{ol97}, and references therein).  From additional optical and
NIR observations, the galaxy appears to contain a Seyfert or
narrow-line AGN.  The intrinsic luminosity, however, is not known due
to possibly significant line-of-sight reddening as the galaxy is seen
nearly edge-on.  Evidence supporting the spiral classification
includes a measured rotation curve from both {\sl O[III]} and {\sl
  H$\alpha$}, a dust-lane bisecting the inner bulge, structure
suggestive of spiral (or ring-like) structure from deep B-band
imaging, surface-brightness profile decomposition showing a very
strong disk component, and a nearly perfect agreement with the
predicted B-band Tully-Fisher relation for a Sa-Sb type galaxy. In
addition to a probable imbedded AGN, the emission-line ratios in the
disk are consistent with typical disk {\sl HII} regions excited by OB
stars.  And interestingly, 0313-192 appears not to be involved in any
type of merging event.

Aside from being intrinsically interesting in itself, 0313-192 offers
a potential testbed for understanding the whole issue of why spirals
generally do not produce powerful jets and double radio sources, which
could have an impact on blackhole, jets, and AGN physics from many
different angles.  While all available evidence would seem to suggest
a spiral classification for the host galaxy and an FR I-like radio
source, \citet{paper1} were unable to confirm the presence of
extended, large-scale jet emission in 0313-192.  With the goal of
looking for both the inferred jets and additional properties which
might explain the peculiar nature of this object, we have reobserved
0313-192 with the VLA at both higher resolution and sensitivity.  In
this paper we report the results of these observations as well as a
search for {\sl HI} and {\sl CO} (sections 2.1-2.3), a radio continuum
spectrum of the nucleus ( section 2.4), and multifiber spectroscopic
observations of the cluster field to understand the properties of the
external environment (section 3).  We discuss the nature of this
source as gleaned from these observations in section 4, and summarize
with our conclusions in section 5.

\section{Radio Observations} 

\subsection{VLA Imaging} 

We reproduce the VLA image and optical (R-band) overlay of 0313-192
from \citet{paper1} and \citet{ol97} in Figure 1.  This image was
produced from VLA snapshots in both the B and C-arrays, with a
resultant beam of $12.1\arcsec \times 9.9\arcsec$ at PA=-1.9$^\circ$.
The two lobes extend $\approx$ 100 $h_{75}^{-1}$ kpc both north and
south of the galaxy.  We measure a total flux density of 98 mJy on
this map, nearly half (48 mJy) originates in the unresolved core.  The
northern radio lobe measures 32 mJy, the southern lobe 18 mJy.  From
both the morphology and the measured radio power ($\log P_{20cm} =
23.95~W~Hz^{-1}$), we classify 0313-192 as a FR I radio galaxy. The
morphology is very similar to the ``fat-double'' class. The host 
galaxy is nearly the faintest radio source in our complete survey of 
Abell clusters \citep{paper6}, and the only one consistent with a
spiral/disk morphology.  While at this resolution we do not detect
actual jet emission, the nature of the radio morphology would
certainly imply the existence of jets as a necessary mechanism for
filling the large-scale lobes.  We can rule out star formation as the
origin of the radio emission based on the radio/FIR relation and
$q$-value \citep{helou85}. \citet{paper1} found $q\leq 0.72$ (as
compared to $<q>=2.27\pm0.20$ for star formation).  The optical and
radio cores align to better than 1$\arcsec$.

We report new VLA observations from 4,5 November 1996 in the A
configuration.  We observed 0313-192 for 4.3 hours at 3.6 cm (X-band)
using the continuum correlator and a 50 MHz bandwidth, and for 4.3
hours at 20cm (L-band) with the spectral-line correlator (4IF mode; 7
channels at 3.125 MHZ with 2 IF's and 2 polarizations), with a total
effective bandwidth of 43.7 MHz over each IF sideband. The $1\sigma$
noise-level on the maps are 13 $\mu$Jy and 24 $\mu$Jy at 3.6 and 20 cm
respectively.  We used the source 0432+416 (8.6 Jy at 20cm) as the
bandpass calibrator. The resultant beams are $2.9\arcsec \times
1.7\arcsec$ at PA=-6.1$^\circ$ at 20cm and $0.3\arcsec \times
0.2\arcsec$ at PA=3.2$^\circ$ at 3.6cm.  We also observed 0313-192 on
11 October 1997 in the hybrid DnC configuration at 20cm with
resolution $\approx$ 28 arcsec.  Total on-source integration was
$\approx$ 2 hours. The observations were once again made in the 4IF
spectral-line mode in order to minimize bandwidth smearing and to
allow for easier removal of interference. The $1\sigma$ RMS noise on
the map is 36 $\mu$Jy. In all cases we used 3C48 as the primary flux
calibrator.  All datasets were CLEANed and self-calibrated using the
AIPS (Astronomical Imaging Processing System) software package.

In Figure 2 we show the low-resolution 20cm image of 0313-192 from the
DnC configuration.  The image is tapered to a beam-size of $46\arcsec
\times 37\arcsec$ in order to bring out the low surface brightness
emission in the lobes.  In comparison to Figure 1, we actually see
that the lobes are likely more extended than previously thought.  From
the map in Figure 2, we estimate a size of $\approx 328h_{75}^{-1}$
kpc (281 arcsec), extending from a declination of $-19^\circ~04'~09''$
to $-19^\circ~08'~55''$.  We measure an integrated flux-density of 106
mJy from this image (as compared to 98 mJy from the map in Fig 1),
corresponding to a radio power of $\log P_{20cm}=23.98$ W Hz$^{-1}$.
At this radio power, 0313-192 is actually substantially bigger (by a
factor of 3) than predicted from the Power-Size diagram for both
cluster and non-cluster radio galaxies \citep{ledlc}. In the
radio-optical luminosity plane, 0313-192 lies exactly on the dividing
line between FR I and FR II sources, and hence at the upper-extreme of
FR I properties for the host luminosity.

We show the A-array high-resolution maps in Figures 3 and 4.  In
Figure 3, we show the 3.6cm image, indicating clear evidence for a
nuclear jet oriented 143$^\circ$ to the SW (measured North to West).
The jet is 1.6\arcsec\/ in length measured from the core to hot-spot,
or $\sim$ 1.9$h_{75}^{-1}$ kpc. At $\sim$ 0.3\arcsec/beam, our
resolution is $\sim$316 pc.  We estimate a flux-density for the jet of
729 $\mu$Jy at 3.6cm corresponding to $\log P_{3.6cm} = 21.82$ W
Hz$^{-1}$.  This jet is nearly identical in length to the M87 jet,
although it is $\approx$ 10 times fainter \citep{owenm87,meis}.  While
nuclear jets are frequently seen in Seyfert galaxy samples (spiral
hosts), the size of these jets are nearly always much smaller than
seen in 0313-192, typically of order pc to hundreds of pc
\citep{kuk99,kuk95,ulv95}. Thus, even on the kpc-scale, 0313-192
appears unlike the typical radio source found in an AGN/spiral-host.

In Figure 4, we show the 20cm A-array image in grey-scale. While the
lobe emission is mostly resolved out, the main feature in this image
is the jet and hot-spots.  Measured from the core to the hot-spot in
the southern lobe, the jet extends 36\arcsec or 42$h_{75}^{-1}$ kpc in
length!  The distance from the core to the nearest hot-spot is
12.7\arcsec\/ or 14.8$h_{75}^{-1}$ kpc. We estimate a total integrated
flux-density of 3.7 mJy (including the hot-spot at the end of the jet
in the southern lobe), or $\log P_{20cm}^{jet} = 22.5$ W Hz$^{-1}$.
If one were to connect back to the nuclear jet seen in the 3.6cm
image, it appears that the jet bends, or changes its position angle
from 143$^\circ$ to 154$^\circ$ at the middle hot-spot to 167$^\circ$
at the hot-spot within the southern lobe. There is another, fainter
knot within the jet at 23\arcsec\/ or $\sim$ 27 kpc from the core.  We
see no evidence for a jet to the north to the limit of our
sensitivity.  The detection of this large-scale jet/lobe morphology
unambiguously confirms our classification of this object as an FR I
radio galaxy, and makes 0313-192 unique in this regard because it is
found in a spiral/disk host galaxy.

\subsection{HI Absorption} 

We obtained an HI aperture synthesis observation of 0313-192 with the
VLA in the D configuration on 24 October 1997.  Only 19 antennas were
available for the observations.  3C 48 was observed as the primary
flux calibrator, and 0237-233 (6.5 Jy) was observed approximately
every 45 minutes for the bandpass calibration as well as for short
term gain and phase calibration. The 64 spectral channels span a 3.125
MHz bandpass centered at $V_{Helio}$ = 20100 \kms, with a spectral
resolution of 48.8 kHz (11.2 \kms channels and a total bandwidth of
717 \kms.).  We observed 0313-192 for 130 minutes, producing a
synthesized beam of $51\arcsec \times 32\arcsec$ at PA=-66$^\circ$.
The RMS noise of the spectrum is 1.2 mJy beam$^{-1}$.  The data were
calibrated and imaged using standard reduction procedures in AIPS.

The HI spectrum is shown in Figure 5 at four separate positions, with
(0,0) being the position of the radio core.  Three other spectra are
included for comparison.  The (+810,+180) spectrum (top) is for a 13.4
mJy unresolved continuum source at an offset position of +810\arcsec\/
in RA and +180\arcsec\/ in DEC. This spectrum demonstrates the
flatness of the bandpass.  The other two offset positions are roughly
along the radio jet, with continuum flux densities of 18.6 mJy
beam$^{-1}$ and 10.0 mJy beam$^{-1}$ for the North and South offsets
respectively. The non-detection of optically thick HI absorption at
the offset positions sets an upper-limit to the size or extent of the
absorber along the direction of the jet (\ie, we are not looking
through an extended halo of HI gas).

The peak absorption occurs at $v = 20106$ \kms, at a depth of
34.0$\pm$1.2 mJy beam$^{-1}$ below the continuum.  Thus, nearly the
entire 20cm core continuum flux ($\approx$ 35.1 mJy - see $\S$2.3) is
absorbed. The FWHM of the line is only 34 \kms\/ with a measured
equivalent width of 32 \kms.  The apparent optical depth for this
narrow absorption feature is then $\tau = 0.98 \pm 0.06$. Since the
peak absorption occurs in a single channel, it seems likely that the
actual $\tau$ is even larger.  We convert the optical depth to an HI
column density from the relation

\begin{equation}
N_{HI} = 1.82\times 10^{18}~T_{S}~\left(\frac{1}{f}\right)~\tau~{\Delta}V~~cm^{-2}
\end{equation}

\noindent 
where $T_S$ is the electron spin temperature, $f$ is the fraction of
the continuum source covered by the absorber, and ${\Delta}V$ is the
velocity width (we use the FWHM = 34 \kms).  The narrow width suggests
that the gas is cold (based on kinematics).  In high density regions,
$T_S$ approaches the kinetic temperature ($T_K$). In the radiation
field of an AGN, irradiated clouds can have a much higher $T_S$, which
may suppress {\sl HI} absorption from an obscuring torus
\citep{gall99}. However, {\sl HI} studies in general are most
sensitive to the coldest regions of clouds which have the highest
densities. With typical narrow-line conditions the high density
results in a collisionally dominated gas which drives $T_S \rightarrow
T_K$. Thus, $T_S \approx 100 K$ appears to hold up under a broad range
of conditions \citep{mht96}.  It is of course possible that the {\sl
  HI} absorption originates in a warmer atomic medium in the AGN
environment, in which case $T_K$ and $T_S$ might approach several
thousand Kelvin before hydrogen is significantly ionized
\citep{gall99}. Thus, $T_S=100$ K provides a lower-limit to the true
{\sl HI} column. With these values we estimate $N_{HI}=6.0\times
10^{21}$ cm$^{-2}$.  One finds nearly the same value considering the
absorption only in a single channel ($\Delta V$=11 \kms, but with an
apparent optical depth of $\tau \geq 3$!). As we are limited in both
spatial and velocity resolution, the observed narrow feature would
suggest that the actual column density is likely much higher ($>
10^{22}~cm^{-2}$).  Note, that this places 0313-192 at the upper
extreme of observed column densities in Seyfert galaxy samples.  The
apparent optical depth is nearly unprecedented.

Optical depths $\geq 1$ are very rarely seen in general, which may
make 0313-192 somewhat unique in yet another respect.  While higher
resolution is clearly needed, these results suggest that the absorbing
gas has to be of high density, and/or we are privileged to have a
sight line through the disk which maximizes the amount of absorbing
gas.  The small line width also suggests we are looking at dynamically
cold gas. The line width in particular places fairly strict limits on
the location of the gas relative to the nucleus because of the depth
of the potential and the radius probed by the absorption.  The HI
absorption properties of 0313-192 bears some resemblance to the model
for MRK 6 from \citet{gall99} -- a disk-like distribution of cold gas
seen against a bright radio core.  The majority of HI absorption
systems associated with AGN hosts show line widths in excess of 50
\kms because the absorption occurs fairly close to the nuclear region,
and also because the line-of-sight passes through a large range in
radius (and corresponding velocity gradient). The much narrower width
we see for 0313-192 would suggest we are not looking at gas close to
the nucleus, but rather gas originating in the ISM.  The {\sl HI}
spectrum of 0313-192 is remarkably similar to that seen for Sgr A$^*$,
${\Delta}V \sim 40-60$ \kms\/ \citep{liszt,falcke98}, despite the
difference in galaxy inclination ($i \sim 80-85^{\circ}$ for
0313-192). With our present spatial resolution it is difficult to know
precisely the origin and location of the absorbing media. The high
optical depth and column density argue that we might be looking
through a spiral arm feature or a warp, that would increase the amount
of line-of-sight absorption over a typical location in the disk viewed
from this angle.  Alternatively, we could be looking through a dense
cloud. \citet{dickey83} find a number of line-of-sight positions
through the galactic plane at $\sim 10^{\circ}$ with equivalently high
column densities.  Thus, we may be looking through the ISM of a fairly
normal spiral galaxy not unlike our own.  If there are any doubts as
to the optical morphology of the host galaxy, these observations would
suggest an additional argument in favor of a spiral host.

\subsection{CO Observations} 

An aperture synthesis CO observations of 0313-192 was carried out with
the Owens Valley Millimeter Array (OVRO) on 26 May, 1997.  There were
six 10.4 m diameter telescopes in the array, providing a field of view
of about 65\arcsec\/ (FWHM) at 108 GHz.  The telescopes are equipped
with SIS receivers cooled to 4 K, and the typical single sideband
system temperature was about 350 K.  Baselines of 15-115 m E-W and
30-80 m N-S were used, and the robust-weighted data after 6 hours of
observations resulted in a synthesized beam of $3.9\arcsec \times
7.9\arcsec$ ($PA=-4^\circ$).  A digital correlator configured with
$120 \times 4$ MHz channels (11.1 \kms) covered a total velocity range
of 1330 \kms.  Uranus ($T_B=120$ K) was observed for the absolute flux
calibration, and nearby quasar 0346$-$279 (0.92 Jy) was observed at 20
minute intervals to track the phase and short term instrument gain.
The data were calibrated using the standard OVRO program {\sl MMA}
\citep{scoville} and were mapped an analyzed using the imaging program
{\sl DIFMAP} \citep{shepherd} and AIPS. The uncertainty in absolute
flux calibration is about 15\%.

A plot of the CO spectrum is shown in Figure 6.  No absorption is seen
at the level of the noise, which is 15 mJy/beam in each of the 8 MHz
(22.2 \kms; covering two 4 MHz channels) channel maps.  This sets a
$3\sigma$ lower limit on the line to continuum ratio of $\leq 0.7$, or
a maximum opacity of 1.2 for the CO line.  Invoking a weak LTE
approximation, an upper limit to the H$_2$ column density can be
derived using Eq.~6 in \citet{wc97}.  If the CO line width is the same
as HI (34 \kms), then a limiting H$_2$ column density of $5\times
10^{19}$ cm$^{-2}$ is derived for $T_s=10$ K and CO abundance of
$10^{-4}$.  This limit is nearly two orders of magnitudes smaller than
the derived HI column density, and the simple inference is that the
absorbing gas is nearly entirely atomic in nature; an unexpected
result for the central region of a disk galaxy.  This discrepancy can
be reduced somewhat if the $T_s$ for the HI and the CO abundance are
both significantly lower than assumed values, but this would be
unusual for the ISM found in the central kpc disk of a late type
galaxy.  Alternatively, a difference in the source-absorber geometry
may offer a plausible explanation, meaning that the atomic and
molecular gas would need to have different spatial distributions.
Perhaps the proximity to the central AGN along a fortuitous line of
sight is giving us a biased sampling of a more extended gas
distribution in the disk.

In order to examine the spectrum for a possible broad CO emission
feature, we summed the spectrum over 16 channels (64 MHz or 178 \kms
per channel).  No CO emission feature is detected at any significant
level, however, and a $3\sigma$ upper limit of M$_{H_2}$ $\leq 4.5
\times 10^9~M_\odot$ is derived assuming a line-width of 355 \kms\ and
using standard CO-H$_2$ conversion.  This limit is comparable to that
of the Milky Way at $M_{H_2} = 3.6 \times 10^9~M_\odot$
\citep{sanders84} and is substantially smaller than what is typically
found among IR luminous, gas-rich mergers ($M_{H_2} \sim
10^{10}~M_\odot$; see Sanders et al. 1991).  These results coupled
with the implied $HI$ column-density and high optical depth are all
suggestive of looking through a Milky-Way like ISM nearly edge-on
against a bright, unresolved, obscured radio core.

\subsection{Core Properties and Nuclear Spectrum} 

As a result of our 3.6cm observations to detect the inner kpc-scale
jet, we also discovered that the core of 0313-192 was significantly
brighter at 3.6cm as compared to 20cm (97 as compared to $\sim$ 37
mJy).  We obtained Target-of-Opportunity observations with the VLA in
early January 1997 in order to determine the core radio spectrum at 6,
2, 1.3, and 0.7 cm ($\sim$5, 15, and 43 GHz).  In Figure 7 we show the
nuclear continuum spectrum of 0313-192. The actual flux-densities are
given in Table 1.  The 6 points to the left are from the VLA.  We also
include two continuum measurements from OVRO (108 and 95 GHz; 2.8 and
3.1 mm). The points to the right include the IRAS detections at both
100 and 60$\mu$m, $3\sigma$ upper-limits at 12 and 25$\mu$m, and
nuclear magnitudes at 1.25 $\mu$m (J) and 2.2 $\mu$m (K)
\citep{paper1}.

Because of the lower resolution of the 20cm observation, the kpc-scale
inner jet (Figure 3) is included within the beam, and will bias the
estimate of the core brightness.  We have estimated the contribution
from the jet at 20cm by scaling the flux-density at 3.6cm to 20cm
assuming a non-thermal spectrum with $\alpha=0.6$ ($F_\nu \propto
\nu^{-\alpha}$).  The peak intensity at 20cm from the map in Figure 4
is 37.3 mJy. The estimated jet contribution is then $\approx$ 2.1 mJy.
We therefore adopt a value of 35.1 mJy for the 20cm core flux density.
As the spectrum turns over somewhere between 6 and 20cm, the shape of
the spectrum is important in this region with regards to possible
interpretations of the spectrum.

First, we consider only the radio core spectrum between 1.4 and 43
GHz.  The spectrum rises very slowly with frequency between 4.9 (6cm)
and 43 GHz (0.7 cm).  A linear fit to the spectrum over this range in
frequency gives a slope of $F_\nu \propto \nu^{0.2}$. An inverted or
nearly flat core spectrum is not uncommon in AGN or blazar like hosts
with bright radio cores, and suggests that we are looking at a very
small central source (like that around a central, supermassive black
hole).  The turnover between 4.9 and 1.4 GHz is mostly likely due to
absorption, either free-free or possibly synchrotron self-absorption.
As our spatial resolution is unsufficient to set a size to the core
region, we model the turnover as due to a free-free absorbing medium
along the line of sight to the compact nucleus.  In this case, the
observed core continuum spectrum should follow:

\begin{equation}
S_\nu = S_0~\nu^{-\alpha}~e^{-\tau_\nu} 
\end{equation}

\noindent
where $S_0$ is the flux density of the nuclear source, $\alpha$ is the
spectral index of the core, and $\tau_{\nu}$ is the frequency
dependent optical depth.  The optical depth is related to the emission
measure and electron temperature via

\begin{equation}
\tau_\nu~=~8.3 \times 10^{-2}~T_e^{-1.35}~\left( \frac{EM}{pc~cm^{-6}}\right)~
\left( \frac{\nu}{GHz}\right)^{-2.1}
\end{equation}

\noindent
where EM is the emission-measure of the gas ($EM=\int n_{HII}^2~dl$),
$T_e$ is the electron temperature, assumed to be $T_e=10^{4}$ K, and
we set $\tau_\nu = \tau_1~\nu_{GHZ}^{-2.1}$.  The result of our fit to
the radio spectrum between 1.4 and 43.3 GHz is shown in Figure 7.  We
find $\alpha=-0.18\pm 0.04$, $\tau_1 = 1.3 \pm 0.2$ and a $1/e$
normalization of $62.7 \pm 6.2$ mJy.  The turnover in the spectrum
between 4.9 and 1.4 GHz can therefore be explained by viewing the
compact core through a free-free absorbing medium with $\tau_{1.4GHz}
= 0.6$ with $EM=4\times 10^6~pc~cm^{-6}$.  If we adopt a galaxy
inclination of 80-85$^\circ$ \citep{led96}, and assuming the
absorbing cloud is located 100 pc from the mid-plane, we estimate a
radius of 500 pc.  From equation (3) we then derive $n_{HII}^2~L =
1.3\times 10^{25}~cm^{-2}$, giving $n_{HII} \sim 89~cm^{-3}$ for
$L=500~pc$.  Once again, these numbers are very similar to what one
would expect looking at a compact source through a typical spiral disk
at fairly high inclination. 

We also include two independent continuum measurements of the core
flux-density near 3 mm obtained with the OVRO.  The continuum
measurement is from the analog correlator with 1 GHz bandwidth.  The
map has an RMS noise of 2.5 mJy/beam.  The continuum source (centered
on the nucleus) is unresolved with a measured integrated flux density
of $64.8 \pm 4.2$ mJy at 108 GHz. We had some concerns about the
absolute flux calibration at OVRO in April and May 1997 due to some
bad weather (in April 1997) and apparent de-correlation on the longer
baselines.  In order to reconfirm our flux-density measurement, we
reobserved 0313-192 at OVRO in the lowest-resolution configuration on
22 April 2000 during excellent weather.  At 95 GHz we confirm a
measurement of $68 \pm 3$ mJy from a total on-source integration of 30
minutes.  Mars was used as the primary flux calibrator. This
measurement agrees well with the May 1997 measurement, and we thus
believe the turnover in the spectrum from 43 to 108 GHz is verified.

What is the nuclear radio-continuum spectrum telling us?  The
low-frequency turnover appears to be well explained by viewing the
central source through a fairly high column-density, free-free
absorbing medium.  With the high-frequency turnover from 43 to 108
GHz, we may be seeing the transition from optically thick radio
emission to optically thin higher frequency emission (an inner size
scale to the optically-thin emission) in the most compact region of
the inner jet. Spectral turnovers in this region of the spectrum are
commonly seen in flat-spectrum radio quasars and blazars
\citep{myun,bloom,gear,mm}. Multi-epoch observations at both 1.4 GHz
and 108 GHz appear to rule out any strong variability in the core
which might also explain the apparent turnover.

While the FIR points are far-less reliable, we can rule out
contamination from other sources within the IRAS beam with some
certainty.  There is only one other galaxy within the IRAS beam, (see
Figure 1) and it is not detected in our radio observations (down to a
$3\sigma$ limit of 72 $\mu$Jy at 20cm or $\log P_{20cm} \leq
20.8~W~Hz^{-1}$).  The apparent excess of FIR emission in 0313-192 is
similar to the FIR/radio spectra of Seyfert galaxies \citep{mm}, where
the submm-FIR emission is mostly thermal emission from dust, heated
either by stars or the AGN.  However, without more complete spectral
information any interpretation is very uncertain.  A higher resolution
FIR observation with SIRTF would enable us to isolate non-thermal and
thermal contributions to the FIR flux.

\section{The Environment around 0313-192}

\subsection{Is Abell 428 a Rich Cluster?} 

As seen in the previous sections, there is clear evidence for
large-scale jet emission in this object, and thus 0313-192 is likely
the first bona-fide spiral/disk galaxy with jets observed on physical
scales much larger than the extent of the host galaxy.  One possible
reason for the apparent rarity of such objects may be due to the
necessity of a confining external medium.  As spiral galaxies tend to
be found in environments with both lower galaxy and gas densities than
ellipticals (the typical host of FR I radio galaxies), one would need
to find a spiral galaxy in an environment rich enough to support an
intragroup/intracluster gaseous medium (IGM/ICM). While
\citet{ponman96} find that spiral-rich groups are very often detected
in X-rays, it is unclear what is the true nature of the emission;
related to an actual diffuse component and external gaseous medium, or
rather originating solely from AGN's and the interstellar medium of
individual galaxies.  Another possibility is that the diffuse gas in
spiral-rich environments is too cool to produce appreciable X-ray
emission \citep{mulchaey2000}.  Spiral-rich groups tend to have lower
velocity dispersions, and correspondingly lower implied virial
temperatures, consistent with this possibility \citep{mulchaey96}.  In
the case of 0313-192, the presence of the large-scale jets and lobes
makes it virtually certain that some kind of medium must be present.

Of particular interest in Abell 428 is the fact that the radio source
and host galaxy are found very near (0.05 corrected Abell radii, or
100$h_{75}^{-1}$ kpc) Abell's catalogued position of the cluster
center.  A survey of the galaxy population within a 45 arcmin field
from the POSS I indicates a very high spiral fraction of $\geq$50\%
based strictly on visual identification of the galaxies.  This result
contrasts strongly with Abell's richness counts ($N_{Abell}=47$),
which is very nearly a Richness Class 1 cluster.  While Abell 428 has
not been the target of a pointed X-ray observation, we are able to set
a upper-limit of $L_X(0.5-2.0~keV)~< 2.4 \times 10^{43}~
h_{75}^{-2}~ergs~sec^{-1}$ \citep{xraypap} within a 500 kpc aperture
from the ROSAT All Sky Survey.  Based on the fit given in
\citet{xraypap}, at this richness one would expect an average $L_X$
approximately 5 times higher than the estimated upper-limit.

In order to understand better the environment around 0313-192, we
observed this cluster with the MX multifiber spectrograph and Steward
Observatory's 2.3-m telescope on the nights of 12-14 December 1999.  A
detailed description of the design of the MX spectrometer is given by
\citet{mx86}.  In brief, the MX spectrograph has 32 fibers available
for object targets and another 30 used to obtain a sky spectrum.
These observations were made with the new 2-arcsec fiber set installed
in 1997.  The Field-of-View (FOV) of the instrument is 45 arcmin.  The
MX fibers are coupled to the Steward Observatory B\&C spectrograph
with a 1200x800 Loral CCD. We used a 400 line mm$^{-1}$ grating
covering the range 3600-6900$\AA$ with a dispersion of
2.75$\AA/pixel$.  We observed Abell 428 with 5 fiber configurations,
targeting 115 different objects within the FOV.  Heliocentric
velocities were determined using the IRAF Fourier cross-correlation
program FXCOR.  For templates we used MX spectra of the galaxies; M31,
M32, NGC 7331, NGC 3379, and the radial velocity standard stars; HD
90861 (K2 III) and HD 4388 (K3 III).  Velocity errors were calculated
from the S/N of the cross-correlation, parameterized by the
\citet{td79} R-value.  We used the relationship $\Delta V$ = 487/(1+R)
\kms, derived from 251 redundant MX observations of cluster galaxies
\citep{pink2000}.  Confirmation of the redshift was checked manually
from the spectra, and we have rejected objects with $\rm R<3$ as
unreliable.

In Table 2 we report 70 (68 new) redshifts (given as $v=cz$) in the
direction of this cluster. The velocity field is shown graphically in
Figure 8a,b.  Clearly there are two primary peaks in the distribution.
Interestingly, there seems to be nearly continuous velocity coverage
from 15000-28000 \kms.  Restricting the objects to velocities in this
range, we have applied the KMM algorithm to objectively partition the
data into subgroups \citep{kmm93,kmm95}.  Briefly, this algorithm
models the velocity distribution as composed of a user-defined number
of subgroups, and applies a mixture-modeling technique to determine
the goodness of fit for a number of Gaussian distributions as compared
to a single distribution.  If multiple groups are detected, the
objects are assigned (with a probability likelihood) to specific group
membership.  We find $>99.99$\% probability that two Gaussian
distributions are a better fit than a single distribution.  Using the
standard biweight estimators of location and scale: $C_{BI}, S_{BI}$
\citep{beers90}, two peaks are found at $C_{BI}^1=20044$ and
$C_{BI}^2=23336$ \kms with scales $S_{BI}^1=347_{-63}^{+89}$ and
$S_{BI}^2=298_{-108}^{+161}$ \kms ($1\sigma$ errors are given on the
dispersions).  The KMM partitioning finds a mixing proportion of 57\%
(24) and 43\% (18) respectively for groups 1 and 2 after applying a
$3\sigma$ clipping to the velocities around each peak.

The measured velocity dispersions for these {\it groups} are nearly
identical to the median dispersion ($295\pm31$ \kms) of a complete
sample of nearby poor galaxy clusters \citep{led96}, suggesting that
0313-192 lives in a much poorer environment than suggested by Abell's
richness estimate.  Using the $L_X-\sigma_V$ scaling-law observed for
poor groups; $L_X \propto \sigma_V^4$ \citep{cfagroups,mz98}, we would
expect an $L_X$ for Group 1 very near to our estimated upper limit.
Thus, it may not be surprising that it was not detected. Based on
these results Abell 428 is really a projection of multiple groups and
not a rich cluster.

\subsection{Radio Source Confinement} 

One of the issues related to finding any extended radio source is
confinement.  In general, it is required that the pressure in the
external environment (the intracluster gas) be comparable to the total
pressure (from fields and particles) in the radio lobes in order to
confine the source.  Otherwise, the radio plasma will simply expand
into some type of equilibrium, with a subsequent large decrease in the
energy density of the magnetic field and equivalently a large drop in
the observed radio luminosity.  The lobes could of course be
overpressured relative to the external medium, but at least some
external, gaseous environment is believed necessary in order to
maintain the luminosity of such large, extended sources. There appears
to be a trend for radio power to increase with the X-ray luminosity
($L_X$) of the external gas for extended FR I radio galaxies,
consistent with this interpretation \citep{neal}.

From the radio map in Figure 1, we have estimated B-fields and total
pressures in the radio lobes assuming equipartition in energy between
particles and fields.  We measure the flux-density over one beam area,
and assume that the depth along the line of sight is symmetric to the
beam projected in the plane of the sky.  We assumed a uniform volume
filling factor, transverse magnetic fields, and a synchrotron spectrum
with upper and lower cutoff energies of 10 MHz and 100 GHz with a
spectral index $\alpha = 0.7$ ($F_\nu \propto \nu^{-\alpha}$).
Measured from several locations within the lobes we estimate total
(field+particle) equipartition pressures of $2-5 \times
10^{-13}~dynes~cm^{-2}$.  These pressures are exactly in the range of
thermal gas pressures ($P_{th}=\rho k T/\mu m_H = nkT$) estimated from
X-rays for poor groups of galaxies \citep{doe}, and a factor of 10
lower than most rich, Abell-class clusters because of the difference
in the central gas densities.  Over the range $L_X =
10^{42-43}~ergs~sec^{-1}$, nearly all poor groups have similar
intracluster temperatures of $1-2$ keV, and as the $L_X/\sigma_v$
relationship flattens substantially over this range in $L_X$, Most
poor groups have fairly similar gaseous environments over a large
range in $\sigma_v$ ($\sigma_v \approx$ 100-400 \kms).  Thus, 0313-192
is completely consistent with living in a poor-group environment,
however a much deeper X-ray observation is now needed to quantify this
any further. 

\subsection{Interpretation of the Velocity Data} 

Abell 428 appears to be a projection of at least two systems along the
line of sight with velocity separation 3302 \kms. If this velocity
difference were purely Hubble-flow, this corresponds to a physical
separation of 41.1$h_{75}^{-1}$ Mpc. The nearly continuous velocity
coverage from 15000-28000 \kms, however, may suggest that we are
seeing some type of supercluster or filament-like structure.

In Figure 9a-d we show contour maps of the adaptively-smoothed galaxy
positions. Figure 9a shows the spatial distribution of all galaxies
which were targeted by our observations. This distribution is
indicative of the actual galaxy distribution within the FOV down to
our magnitude limit ($\approx m_R=18.5$), and justifies Abell's
classification of this system as a rich cluster.  In Fig 9c,d we show
the spatial distributions of groups 1 (containing the radio source)
and 2, with the location of 0313-192 marked by a cross.  While group 1
is apparently richer than Group 2, it is also more compact.  The peak
galaxy surface-density in group 1 is nearly a factor of 6.5 times
higher than group 2.  The radio source ($v_H=20143$ \kms) is offset
99 \kms from the velocity peak and 461 kpc in projection to the SW
of the peak in the galaxy surface-density distribution for this group
(group 1).  We see that the distributions are quite different in both
orientation and location.  In Fig 9b we show the spatial distribution
of all galaxies within the 15000-28000 \kms range in velocity which
are not within $1S_{BI}$ of one of the two groups. While the peak is
offset slightly to the west of group 1, it appears that the objects
with velocities in this range are essentially randomly distributed in
the field; \ie a ratio of 9a and 9b would produce something
that is nearly uniform over the areas in common. While we
unfortunately do not have complete velocity information for a
magnitude-limited sample, these observations suggest that we may be
seeing a large complex of galaxy groups and {\it field} along our line
of sight which is suggestive of a super structure or filament $>$40
Mpc in size viewed nearly end-on.

We have also examined the velocity data sets for evidence of
substructures, possibly related to merging, ongoing dynamical
activity, or otherwise unrelaxed conditions.  As tests of the entire
data set would undoubtedly find substructure because there is more
than one distinct group of galaxies both spatially and along our line
of sight (viewed in projection), we consider each group separately for
the tests.  We applied a number of statistical tests of the 1D
(velocity), 2D (spatial distribution), and 3D (velocity + spatial)
distributions, which are summarized in \citet{pinkneystat}.

We find no evidence of 1D substructure for either galaxy group,
consistent with the velocity distributions being essentially Gaussian
However, for Group 1 (which contains the radio source), we do find
statistically significant ($\geq 97\%$) signs of substructure in both
2D and 3D tests.  For the 2D tests, both the $\beta$ and $Lee2D$ tests
were significant (the $\beta$-test at the $>99.99\%$ level).  The
$\beta$-test is a {\it symmetry} test, in that it is sensitive to
significant deviations from mirror symmetry about the cluster center.
It is not sensitive to circular asymmetries, so the elongation of
Group 1 seen in Figure 9 does not explain the result. We interpret the
result as caused by clumpiness in the spatial distribution; there are
significant clumps of galaxies both to the NE and SW of 0313-192,
while 0313-192 has only very few galaxies within a few hundred kpc.
For the 3D tests, the \citet{ds} (DS or $\Delta$-test) was significant
at $> 99.99\%$ for group 1 only.  The DS-test looks for deviations in
the local mean velocity and dispersion as compared to the global mean
and dispersion.  For group 1, quite significant substructure is seen
both to the NE and SW (the most significant) of 0313-192, consistent
with the 2D-clumping found from the $\beta$-test. Note that this does
not necessarily have to be the case; the 2D and 3D tests find
substructure in the same parts of the cluster only because the clumps
in 2D space have low velocity dispersions, and are significantly
different from the mean of the group.  This picture supports the
interpretation that the galaxies within Group 1 are not dynamically
relaxed.  The fact that the substructures are seen on the outskirts of
the group, both spatially and in velocity space, would suggest that
the group is in the process of collapsing or merging-together.  The
lack of 1D substructure, the low velocity dispersion, and the high
compactness of Group 1 are all consistent with this happening more or
less along our line of sight.  Interestingly, this result is similar
to our view of the larger-scale structure being some type of filament
viewed end-on.  While it is not clear to what extent these results are
important in terms of understanding the unusual radio source in
0313-192, we note that the position angle of the radio source appears
to point within $\leq 10^\circ$ of the direction towards the primary
(sub)clump of galaxies to the NE and SW, possibly aligning along a
merger axis, if this interpretation is correct.

\section{Discussion}

We have shown that 0313-192 is an unusual, and in many respects an
unique object.  One then wonders as to why only one such radio source
has so far been detected in a spiral galaxy. Is there something unique
about the host galaxy and the nuclear environment which has allowed a
powerful FR I radio source to form and develop?  Is there something
unique about the external environment around 0313-192?  Is it possible
that while rare, there is a population of such objects which have
simply been missed in existing surveys?  This last point certainly
seems plausible.  In terms of dedicated searches/surveys of radio
galaxies, rich clusters have been much more thoroughly studied than
groups or the field, and spiral galaxies are not terribly common in
the cores of rich clusters.  As we have shown, Abell 428 should not
really be catalogued as a rich cluster at all.  It is possible that
similar objects might be found from the FIRST and NVSS surveys,
although clearly this would require optical identifications and
morphologies for a very large number of objects. These surveys are
also limited, in that while NVSS has the sensitivity to find such
objects, the resolution is poor and only a very nearby object would
appear significantly extended to warrant further inspection.  The
FIRST survey has the resolution, however the surface brightness
sensitivity is much lower, and the extended emission might likely be
missed or resolved out.  Thus, a targeted radio survey of poor groups,
and in particular, spiral-rich groups might gleam more similar
objects.

Our probes of the extra-nuclear environment as seen in {\sl HI}
absorption and a limit on {\sl CO} and the gas mass both strongly
suggest that we are looking through an interstellar medium resembling
a typical spiral galaxy like the Milky Way.  With regards to the {\sl
  HI}-absorption, the optical-depths and column-densities inferred are
consistent with this perspective.  However, the upper limit to the CO
emission is somewhat surprising. The inferred gas mass not only
precludes 0313-192 resembling a typical gas-rich merger remnant, but
the limit is sufficiently stringent to require either spatial
segregation of the two phases in the absorbing medium or an ISM
unusually deficient in molecular gas, which would be unexpected for a
the inner region of a spiral galaxy.  It is possible, though unlikely
(?) that our line of sight to the nucleus intercepted a particularly
high density cloud or region of the inner gas distribution (meaning
that we overestimate the inferred gas mass based on this line of
sight), although the extremely narrow line width is difficult to
understand in this context. Thus, there are still several unanswered
questions, although the general picture of a spiral disk and ISM would
appear to be supported by the observations.

In terms of the nuclear properties, 0313-192 shares many similarities
with QSO or Blazar-like AGN's.  The radio luminosity of the core and
flat continuum spectrum in particular is very similar to the typical
QSO.  However, there are no other objects in nearby Quasar samples
with similar radio morphology.  \citet{bahcall97} studied 20 nearby
QSO's, 3 with spiral morphology.  Two of these (PG 0052+251 and PG
1402+261) are not detected by NVSS or FIRST.  The other source, PG
1309+355 (z=0.184), has a flat-spectrum core (40 mJy at 4.9 GHz and
44 mJy at 1.4 GHz) five times stronger than 0313-192.  However, there
is no evidence for more extended emission or a jet.
\citet{boyce98,boyce99} studied 19 nearby QSO's, 3 with spiral hosts.
Only one of the three (MS 07546+3928, z=0.096) was detected in the
radio.  MS 07546+3928 appears to be resolved on the NVSS, with a total
flux-density of 32.4 mJy at 1.4 GHz ($\log P_{1.4}=23.8~ W~Hz^{-1}$).
On the FIRST image, however, the 32.4 mJy NVSS source is resolved into
two sources (21.2 and 11.2 mJy), 39 arcsec in separation. The weaker
of these two is centered at the position of the QSO.  There is no
evidence from the FIRST image of any diffuse emission connecting the
two sources.  Thus, of the 39 nearby QSO's studied in these surveys, 6
(15\%) have clear spiral hosts.  Among these, only two (5\%) are
detected in current radio surveys.  This is similar to the expected
radio-loud fraction of optically-selected luminous AGN's.

There are some similarities between 0313-192 and IIIZw2, claimed to be
the first superluminal jet found in a spiral/Seyfert galaxy.
\citet{brunthaler} model IIIZw2 as an example of a {\it frustrated}
jet interacting with the dense ISM on the subpc scale.  And, like
0313-192, IIIZw2 exhibits an inverted core spectrum at low frequencies
and a peak and similarly a high frequency turnover somewhere between
0.7 and 3mm (43-108 GHz) range. The radio core in IIIZw2 is however
much brighter and highly variable, brightening by a factor of 20
within only a 2 year time span \citep{falcke99}.  \citet{falcke99}
model the outburst spectrum and timescale as being due to shocks and
very compact hotspots.  Thus, the central AGN in IIIZw2 must be
instrinsically much more luminous than in 0313-192 and the
extranuclear environments possibly very different in order for the
jets to escape to $\gg$kpc scales in the latter.  While there is a
suggestion of a lobe-like structure 15\arcsec to the SW in IIIZw2 (22
kpc at z=0.089), it is not clear that there is a direct connection to
the subpc scale jet and the galaxy itself \citep{kuk98}. If they are
related, IIIZw2 resides at the upper envelope of radio source sizes
for Seyfert galaxies, but is still a factor of 10 smaller than
0313-192. 

This comparison would suggest that 0313-192 is host to a QSO-like
central engine, but that in itself is not sufficient to produce
extended jet and lobe emission typical of FR I radio galaxies.  Thus,
the extended nature of 0313-192 might still be related to a set of
special circumstances.  As pc-scale jets are frequently seen, even in
Seyfert samples (with significantly lower core fluxes than in 0313),
the presence of such a large-scale jet would suggest that both the
initial conditions in the jet and the immediate environment around it
might be somehow different in 0313-192.  0313-192 is clearly at the
upper-end of the radio luminosity function of typical Seyferts, so
should be considered an extreme case in that regard. Exactly how rare
is 0313-192 is not yet known, and more dedicated search will be
necessary in order to find for more examples. Further high-resolution
observations are warranted and will be necessary to probe the
environment within the inner kpc and to address these and other
questions.  Understanding this object may therefore shed light on the
different natures of nuclear activity and perhaps more fundamentally,
will impact our understanding of AGN unification models in general.

\section{Conclusions}

From the new observations presented in this paper we make the
following conclusions regarding the nature of the unusual object
0313-192:

\begin{enumerate}
  
\item 0313-192 appears to be the first confirmed disk/spiral galaxy
  with $\gg$kpc -scale radio jets and double lobed radio morphology.
  The jet would appear to be nearly continuous from sub-kpc to the full
  extent of $\approx 40$ kpc feeding into the southern lobe.

\item The shape of the nuclear radio spectrum strongly resembles a
  QSO or Blazar-like object, rising towards higher frequencies
  (inverted).

\item The nuclear continuum source is nearly completely absorbed by
  {\sl HI} along our line of sight.  Essentially all of the absorption
  falls within a single channel at our resolution (11 \kms, EW of the
  line is 32 \kms) and at the systemic velocity of the galaxy.  The
  implied optical depth is $\tau \geq 1$, which is nearly
  unprecedented and is more extreme than any other observed Seyfert
  galaxy.  We can rule out an extended halo of {\sl HI}, so the {\sl
    HI} much be concentrated much nearer the nucleus, likely in a
  disk-like distribution. The inferred column density is $\approx
  6\times 10^{21}~cm^{-2}$.
  
\item We place a fairly stringent limit to {\sl CO} emission, which
  gives an upper-limit to the molecular gas mass nearly two orders of
  magnitude less than the derived {\sl HI} column density.  So, either
  the absorbing gas is nearly entirely atomic, or the two phases of
  the absorbing material are spatially segregated in this part of the
  disk (along our line of sight and in close proximity to the
  nucleus). Both interpretations are somewhat difficult to understand
  for the inner region of a disk galaxy.  The limit to the gas mass
  based on standard $H_{2}/CO$ conversions is approximately that of
  the Milky Way.  We can therefore rule out with some confidence that
  0313-192 is typical of IR Luminous, Gas-Rich merger remnants, and
  resembles more a typical spiral seen nearly edge-on. 
  
\item We model a low-frequency turnover (4.8-1.4 GHz) in the radio
  continuum spectrum as seeing the source through a fairly high
  column-density ($\approx 10^{25}~cm^{-2}$) Free-Free absorbing
  medium. A possible turnover at higher frequencies (108 GHz) may
  indicate that we are seeing the transition between optically
  thick/thin emission from the most compact region of the inner jet.
  
\item Abell 428 should be reclassified, as the velocity distribution
  of galaxies in the field can be characterized by at least 2 groups
  in velocity space which overlap in their spatial distributions.  A
  nearly continuous range of observed velocities over 13000 \kms
  coupled with the spatial distributions of the clumps suggests that
  we are looking down a fairly extensive filament-like structure of
  order 40 Mpc in extent.  0313-192 lives in the richer and more
  compact of the two groups, although does not appear to reside at the
  dynamical center of its parent group. 

\end{enumerate}

\acknowledgments

M.L. acknowledges partial financial support from NASA grant NAG5-6739.
We thank the referee, S. Antonucci, for posing several questions which
have improved the discussion in the paper.

%% Generally speaking, only the figure captions, and not the figures
%% themselves, are included in electronic manuscript submissions.
%% Use \figcaption to format your figure captions. They should begin on a
%% new page.

\clearpage

%% No more than seven \figcaption commands are allowed per page,
%% so if you have more than seven captions, insert a \clearpage

%% There must be a \figcaption command for each legend. Key the text of the
%% legend and the optional \label in curly braces. If you wish, you may
%% include the name of the corresponding figure file in square brackets.
%% The label is for identification purposes only. It will not insert the
%% figures themselves into the document.
%% If you want to include your art in the paper, use \plotone.
%% Refer to the on-line documentation for details.

%\centerline{\bf FIGURE CAPTIONS}

\begin{figure}
\figurenum{1}
\epsscale{1}
\plotone{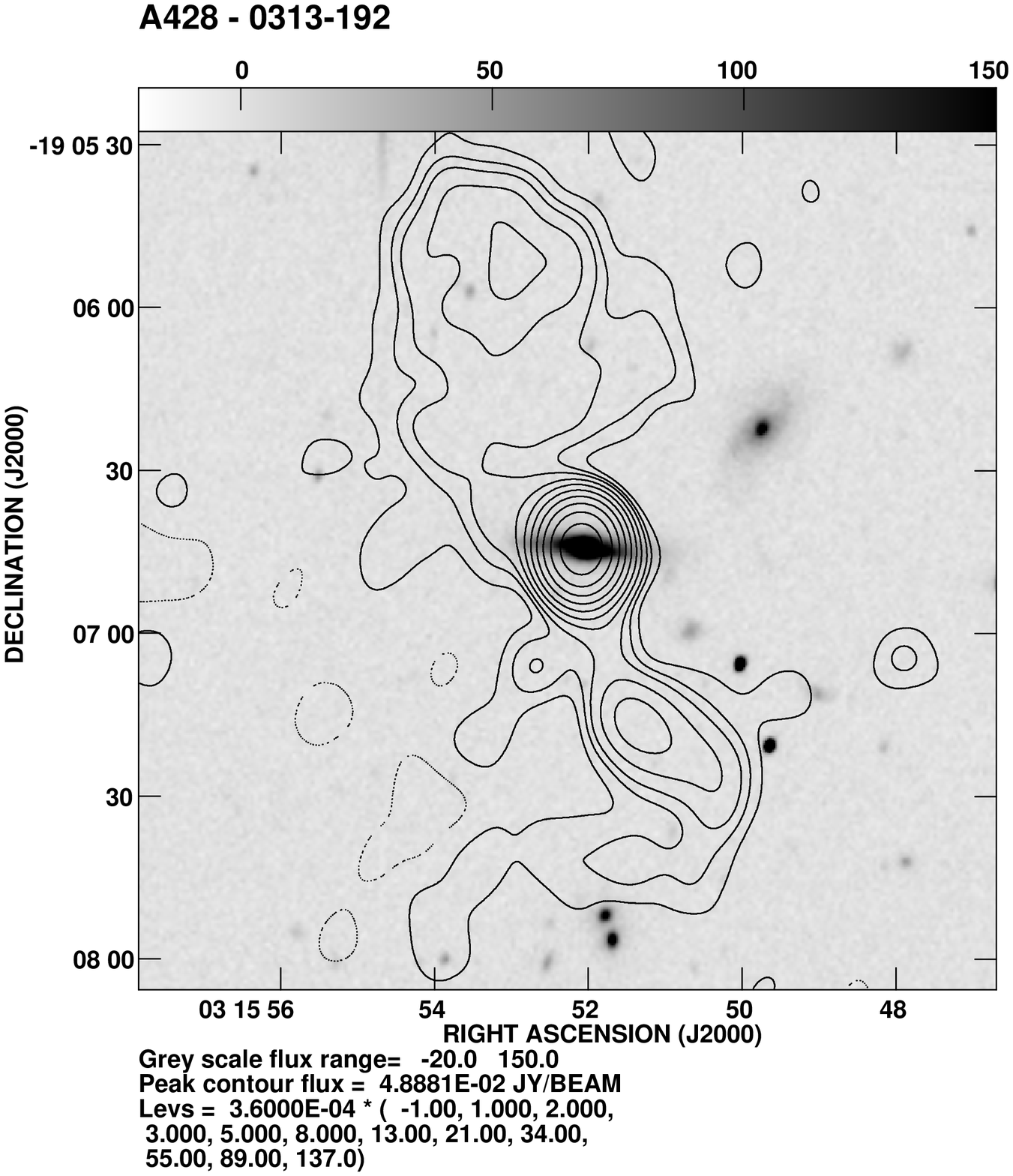}
\caption
%\figcaption[Fig1.eps]
  {Radio (20cm)/Optical (R-band) overlay for 0313-192. The beam-size is
  $12.1\arcsec \times 9.9\arcsec$ at a position angle of -1.9$^\circ$.
  The RMS noise on the radio map is $\approx$ 0.1 mJy. From
  \citet{paper1}.}
\clearpage
\end{figure}

\begin{figure}
\figurenum{2}
\epsscale{1}
\plotone{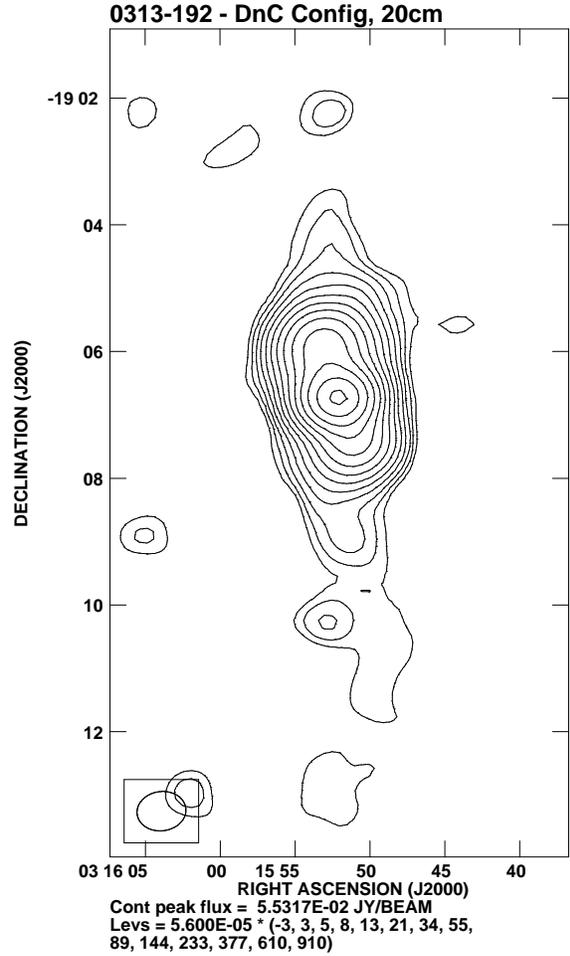}
\caption
%\figcaption[Fig2.eps]
  {Low-resolution 20cm image of 0313-192 from the DnC configuration.
  This image is a tapered map with a resolution of $46.4\arcsec \times
  37.1\arcsec$ at $\theta=-82.6^\circ$. We estimate the source extent
  at 328$h_{75}^{-1}$ kpc from this map.  The RMS noise is 34.8 mJy
  beam$^{-1}$.}
\clearpage
\end{figure}

\begin{figure}
\figurenum{3}
\epsscale{1}
\plotone{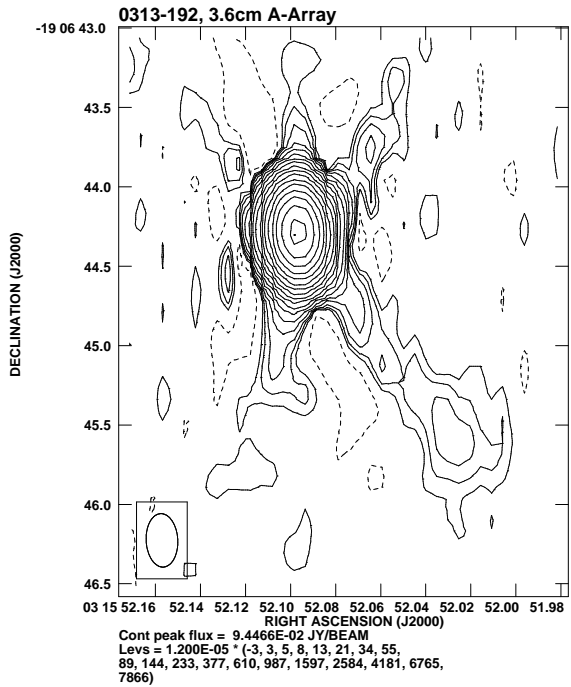}
\caption
%\figcaption[Fig3.eps]
  {High-resolution 3.6cm image made in the A-array.  The beam-size is
  $0.34\arcsec \times 0.19\arcsec$ at $\theta=3.2^\circ$.  The
  $\approx$ 1.6\arcsec (or 1.9 kpc) extension to the SW is the nuclear
  jet.  The other features surrounding the core are artifacts of the
  deconvolution.}
\clearpage
\end{figure}

\begin{figure}
\figurenum{4}
\epsscale{1}
\plotone{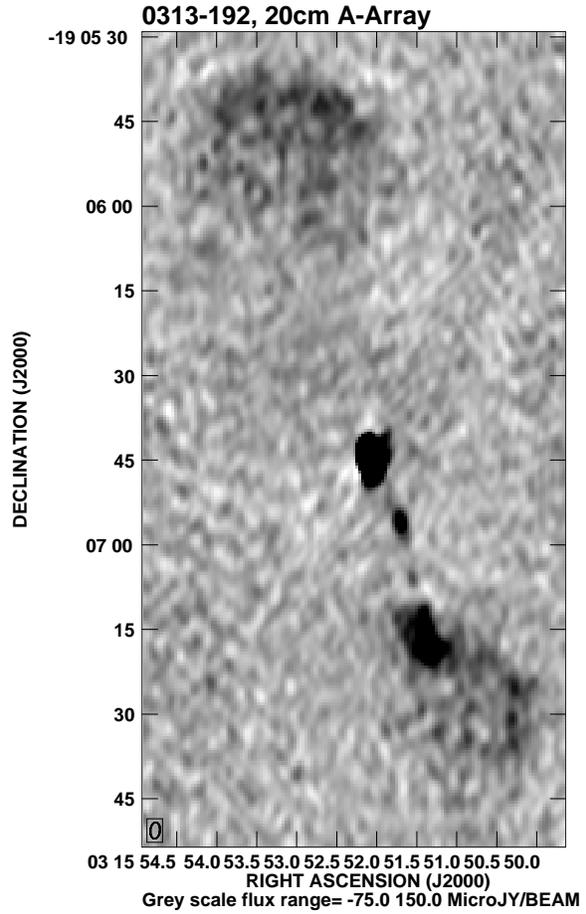}
\caption
%\figcaption[Fig4.eps]
  {A-array 20cm image of 0313-192 at a resolution of $2.89\arcsec \times
  1.69\arcsec$ at $theta=-6.1^\circ$.  We detect a jet with core to
  outer hot-spot length of $\approx$ 41 kpc.  Note the multiple knots
  along the jet, and an apparent change in the position angle of
  $\approx$ 24$^\circ$ from the nuclear jet in Figure 3 to the
  large-scale jet seen here.}
\clearpage
\end{figure}

\begin{figure}
\figurenum{5}
\epsscale{1}
\plotone{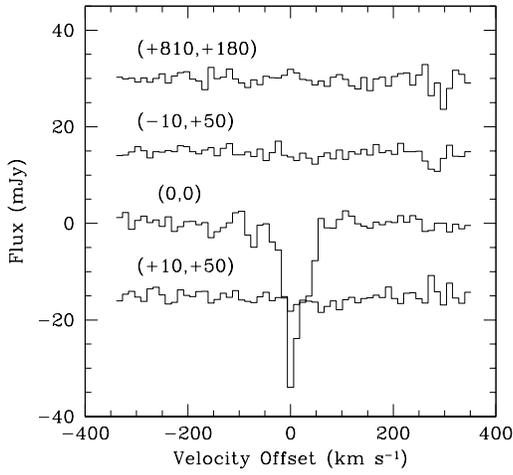}
\caption
%\figcaption[Fig5.eps]
  {HI spectrum for 0313-192.  The top spectrum is that of an unresolved
  13.4 mJy continuum source offset +810\arcsec and +180\arcsec in RA
  and DEC respectively, and shows the flatness of the bandpass.  The
  other spectrum are from positions offset along the radio jet.  HI
  absorption is only seen against the nucleus, and is very narrow;
  FWHM = 34 \kms, ruling out a large, diffuse HI halo.}
\clearpage
\end{figure}

\begin{figure}
\figurenum{6}
\epsscale{1}
\plotone{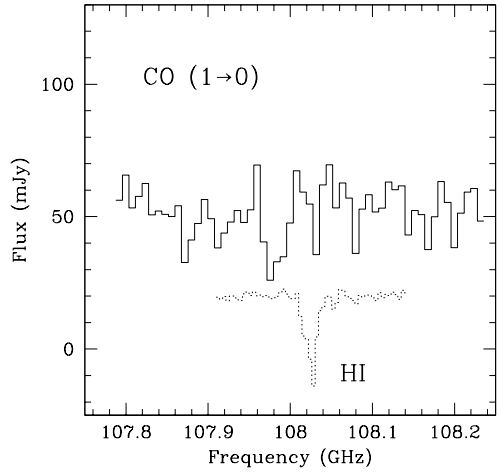}
\caption
%\figcaption[Fig6.eps]
  {CO $1\rightarrow 0$ spectrum of 0313-192 from OVRO.  No absorption is
  seen at the level of the noise, which is 15 mJy/beam in each of the
  8 MHz (22.2 \kms; covering two 4 MHz channels) channel maps.}
\clearpage
\end{figure}

\begin{figure}
\figurenum{7}
\epsscale{1}
\plotone{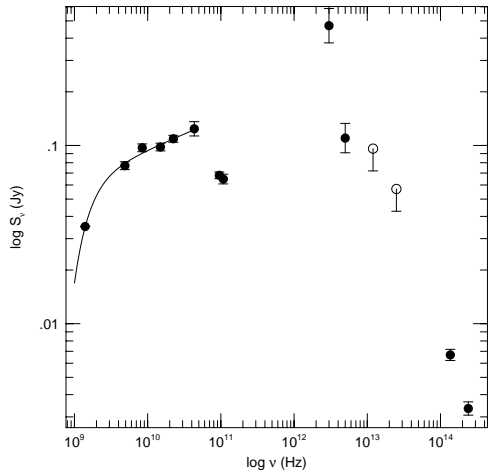}
\caption
%\figcaption[Fig7.eps]
  {Continuum core spectrum for 0313-192. Data points are listed in Table
  1.  The 1.4-43.3 GHz points are from the VLA.  Two points near 108
  GHz are from OVRO. We also show IRAS detections at 100 and 60$\mu$m,
  $3\sigma$ upper-limits at 25 and 12$\mu$m, and J and K-band nuclear
  magnitudes from \citep{led96}.}
\clearpage
\end{figure}

\begin{figure}
\figurenum{8}
\epsscale{1}
\plotone{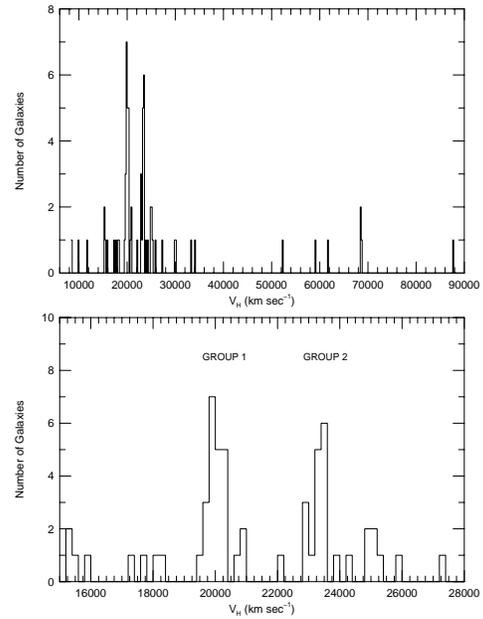}
\caption
%\figcaption[Fig8.eps]
  {Velocity histograms for the Abell 428 field. (A) top: velocity data
  for the 70 measured redshifts in Table 2. (B) bottom: velocities in
  the range $15000 \leq v_H \leq 28000$ \kms.  There are two primary
  groups separated by $\approx 3300$ \kms.  0313-192 is at a
  velocity of 20143 \kms.}
\clearpage
\end{figure}

\begin{figure}
\figurenum{9}
\epsscale{1}
\plotone{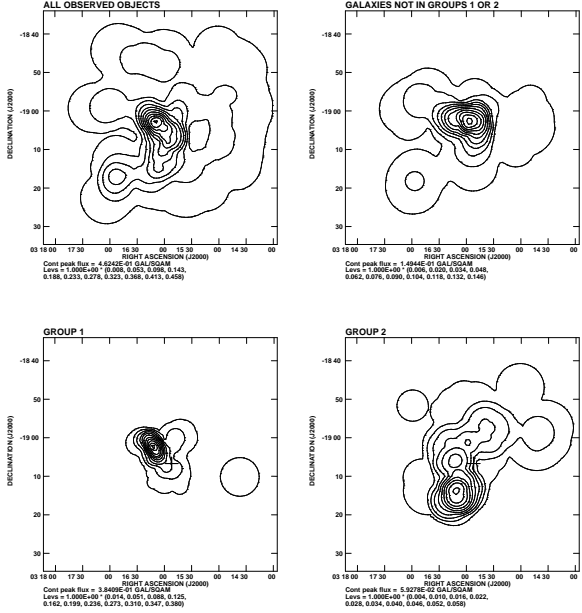}
\caption
%\figcaption[Fig9.eps] 
  {Adaptively-smoothed contour maps of the optical galaxy positions.
  Contour levels are in units of $Galaxies~arcmin^{-2}$. (A) top-left:
  All objects targeted by our observations.  This distribution should
  be reflective of the true galaxy distribution within this field to
  $\approx m_R=18.5$. (B) top-right: All galaxies with measured
  velocities between 15000-28000 which are not within $1\sigma$ of the
  group 1 and 2 velocities.  This distribution represents the $\sim$
  continuous velocity continuity seen in Fig 8b. (C) bottom-left: the
  spatial distribution for Group 1 (which includes 0313-192). The
  cross marks the location of 0313-192.  (D) bottom-right: the spatial
  distribution of Group 2.}
\clearpage
\end{figure}

\clearpage

\begin{deluxetable}{ccc}
\tablecolumns{2}
\tablewidth{0pc}
\tablenum{1}
\tablecaption{Core Spectrum} 
\tablehead{
\colhead{Frequency} & \colhead{Wavelength} & \colhead{Flux-Density} \\
\colhead{(GHz)} & \colhead{} & \colhead{(mJy)}}
\startdata
1.4\tablenotemark{a} & 20 cm & 35.1$\pm$0.1 \\
4.9 & 6 cm & 77$\pm$4 \\
8.5 & 3.6 cm & 97$\pm$5 \\
14.9 & 2 cm & 98$\pm$5 \\
22.5 & 1.3 cm & 109$\pm$5 \\
43.3 & 0.7 cm & 124$\pm$12 \\
95.3 & 3.2 mm & 68$\pm$3 \\ 
108.0 & 3.6 mm & 64.8$\pm$4.2 \\
\nodata & 100 $\mu$m & 470$\pm$117 \\
\nodata & 60 $\mu$m & 110$\pm$23 \\
\nodata & 25 $\mu$m & $<$96 \\ 
\nodata & 12 $\mu$m & $<$57 \\ 
\nodata & 2.2 $\mu$m & 6.7$\pm$0.5 \\ 
\nodata & 1.2 $\mu$m & 3.3$\pm$0.3 \\
\enddata
\tablenotetext{a}{We have subtracted jet-emission on the sub-arcsec
  scale from the 20 cm core-flux density assuming $\alpha=0.6$ ($F_\nu
  \propto \nu^{-\alpha}$), and extrapolating from the 8.4 GHz
  integrated flux-density of the jet to 1.4 GHz.}
\end{deluxetable} 

\clearpage 

\begin{deluxetable}{cccc}
  
\tablecolumns{4} \tablewidth{0pc}
\tablenum{2}
\tablecaption{Velocity Data for A428} 
\tablehead{ \colhead{RA (2000)} & \colhead{DEC (2000)} &
\colhead{$V_H$} & \colhead{$\Delta V_H$} \\
\colhead{} & \colhead{} & \colhead{(\kms)} & \colhead{(\kms)}}
\startdata
  03:15:52.1 & --19:06:45 & 20143\tablenotemark{a} & 66 \\
  03:15:49.8 & --19:06:23 & 15225 & 87 \\
  03:15:49.0 & --19:05:05 & 15266 & 82 \\
  03:16:07.2 & --19:03:35 & 19836 & 80 \\
  03:16:12.5 & --19:04:55 & 19788 & 35 \\
  03:16:16.6 & --19:04:59 & 20318\tablenotemark{b} & 37 \\
  03:16:11.6 & --19:03:06 & 19876 & 31 \\
  03:16:20.8 & --19:00:52 & 20314 & 66 \\
  03:16:23.3 & --19:00:39 & 19502 & 106 \\
  03:16:01.0 & --19:14:35 & 23951 & 98 \\
  03:15:22.2 & --18:59:15 & 9914\tablenotemark{c} & 25 \\
  03:15:50.7 & --19:07:00 & 23584 & 105 \\
  03:16:00.3 & --19:10:08 & 61631 & 100 \\
  03:16:45.1 & --18:56:12 & 15997 & 92 \\
  03:16:40.6 & --18:57:04 & 15599 & 90 \\
  03:16:30.0 & --19:01:42 & 20015 & 79 \\
  03:16:15.1 & --19:00:48 & 19742 & 49 \\
  03:16:07.1 & --19:02:26 & 30175 & 88 \\
  03:16:12.3 & --19:02:40 & 20752 & 107 \\
  03:16:18.2 & --19:02:44 & 22893 & 99 \\
  03:16:04.2 & --19:04:53 & 19948 & 102 \\
  03:15:54.7 & --19:00:56 & 20888 & 89 \\
  03:15:49.4 & --19:11:18 & 20293 & 77 \\
  03:16:06.8 & --19:14:22 & 23447 & 38 \\
  03:16:15.2 & --19:08:55 & 23487 & 84 \\
  03:16:19.0 & --19:04:05 & 22088 & 108 \\
  03:15:48.5 & --19:02:57 & 24943 & 98 \\
  03:15:38.3 & --19:00:21 & 19624 & 27 \\
  03:16:02.8 & --19:14:35 & 34159 & 84 \\
  03:16:44.2 & --19:14:06 & 23409 & 73 \\
  03:16:06.1 & --19:07:25 & 20048 & 115 \\
  03:16:10.6 & --19:09:39 & 20277 & 110 \\
  03:16:14.4 & --19:01:06 & 19929 & 84 \\
  03:15:49.3 & --18:52:12 & 20074 & 112 \\
  03:15:52.2 & --18:57:05 & 25263 & 51 \\
  03:15:34.9 & --18:57:54 & 23148 & 55 \\
  03:15:39.2 & --18:58:38 & 23290 & 100 \\
  03:16:35.8 & --18:45:31 & 19859 & 100 \\
  03:16:00.8 & --18:54:35 & 23276 & 43 \\
  03:17:10.4 & --18:59:36 & 23258 & 55 \\
  03:17:03.4 & --18:59:46 & 18383 & 34 \\
  03:17:19.0 & --19:00:09 & 25081 & 98 \\
  03:16:50.3 & --18:44:28 & 22816 & 51 \\
  03:16:58.0 & --19:18:32 & 27307 & 80 \\
  03:16:57.4 & --19:19:56 & 24872 & 65 \\
  03:16:56.9 & --19:17:11 & 33331 & 75 \\
  03:16:53.8 & --19:18:32 & 87649 & 92 \\
  03:16:24.0 & --19:16:18 & 8406\tablenotemark{c} & 20 \\
  03:16:20.6 & --19:16:24 & 23269 & 66 \\
  03:16:15.0 & --19:17:32 & 23392 & 60 \\
  03:16:21.0 & --19:19:15 & 20283 & 79 \\
  03:16:57.9 & --19:12:04 & 59135 & 65 \\
  03:16:51.5 & --19:12:58 & 17342 & 89 \\
  03:17:20.5 & --19:27:09 & 25017 & 54 \\
  03:15:34.2 & --19:15:07 & 25821\tablenotemark{c} & 50 \\
  03:15:25.1 & --19:19:24 & 19935 & 94 \\
  03:15:04.0 & --19:22:20 & 68600 & 110 \\
  03:15:44.9 & --19:19:58 & 68423 & 58 \\
  03:15:09.6 & --19:11:12 & 52260 & 83 \\
  03:14:46.3 & --19:13:03 & 20042 & 34 \\
  03:14:25.6 & --19:10:49 & 20818 & 97 \\
  03:14:43.3 & --19:08:04 & 19879 & 76 \\
  03:15:04.7 & --19:05:40 & 17600 & 100 \\
  03:14:56.6 & --19:01:03 & 22824 & 77 \\
  03:14:31.6 & --19:00:56 & 23411 & 70 \\
  03:14:38.3 & --18:50:12 & 24372 & 118 \\
  03:14:46.1 & --18:45:24 & 29901 & 70 \\
  03:15:26.5 & --18:44:26 & 23470 & 61 \\
  03:15:52.7 & --19:02:18 & 18069 & 86 \\
\enddata
\tablenotetext{a}{This object is the subject of this paper,
 0313-192. Redshift was also published in \citet{olk}.}
\tablenotetext{b}{This object is the other radio galaxy
 identified in A428 \citep{ol97}. Redshift was also published in
 \citet{olk}.}  
\tablenotetext{c}{Emission-line velocity.}
\end{deluxetable}

\end{document}